\documentstyle[12pt]{article}

\def\pe {p^{(n)}}
\def\G {(\pi^2\;-\;e\;\tilde F \cdot S)}

\begin{document}
\begin{center}

\large {{\bf Comment on ``Anyon in an External Electromagnetic Field:
Hamiltonian and Lagrangian Formulations''}}

\end{center}

\centerline {D. Dalmazi and A. de Souza Dutra }
\bigskip
\centerline {UNESP - Campus de Guaratinguet\'a - DFQ - CEP 12500}
\centerline {Av. Dr. Ariberto Pereira da Cunha, 333 - Guaratinguet\'a -
SP - Brazil}
\bigskip

In a recent paper [1] Chaichian et al. proposed a model for
a free relativistic particle with fractional spin. They have further
studied the interaction of this particle with an external electromagnetic
field. Using their basic brackets (formula (3) of [1]) it is possible
to verify the usual 2+1 spin algebra $\,\lbrace S_{\mu},S_{\nu}\rbrace =
\epsilon_{\mu\nu\lambda}S^{\lambda}\,$. However, after the elimination
of the second class constraints (5b) and (5c) of [1] through Dirac brackets,
one obtains

\begin{equation}\left\lbrace
S_{\mu}\;,\;S_{\nu}\right\rbrace_{DB}\;=\;0,\end{equation}
which shows that the quantities $\,S_{\mu}\,$ cannot be interpreted
as the particle's spin. Consequently the model
suggested in [1] does not describe a free relativistic anyon. Actually,
there is no clear separation in [1] between orbital and intrinsic
angular momenta, this can be seen from the commutation relations
given in formula (11) of [1] and the nonvanishing crossed commutators:

\begin{equation}\left\lbrack x_{\mu}\;,\;n^{\beta}\right\rbrack\;=\;i
\;{\frac
{n_{\mu}p^{\beta}}{p^2}}\;,\;\left\lbrack x_{\mu}\;,\;p^{(n)\beta} \right
\rbrack\;=\;i\;{\frac {\pe_{\mu}p^{\beta}}{p^2}}.\end{equation}

Indeed, since $\,[S_{\mu},S_{\nu}]=0\,$ the crossed commutators (2) are
needed in order to check the closure of the Lorentz algebra
$\,[J_{\mu},J_{\nu}] = i\epsilon_{\mu\nu\lambda}J^{\lambda}\,$ where
$\, J_{\mu}= \epsilon_{\mu\alpha\beta}(p^{\alpha}x^{\beta} +
p^{(n)\alpha}n^{\beta})\,$ stands for the total angular momentum vector.
At this point one might try to redefine the operator $\,S_{\mu}\,$
in order to get the correct spin algebra. Nevertheless, the existence of
the first class constraint
$\,\Phi_2 = S\cdot p + \alpha m \approx 0\,$ restricts
this redefinition to terms proportional to the vectors $\,n_{\mu}\,$
and $\,\pe_{\mu}\,$, but such redefinition must be discarded since
it would lead to the nonconservation of the new spin operators.

Still following [1], analogous problems appear when we turn on the
external electromagnetic field. Once again we get the correct spin
algebra before the elimination of the second class constraints
(18b) and (18c) of ref.[1], but after that we obtain:

\begin{equation}\left\lbrace S_{\mu}\;,\;S_{\nu}\right\rbrace_{DB}\;=\;
{\frac {-\;e\;\tilde F \cdot S}{\G}}\;
\epsilon_{\mu\nu\lambda}S^{\lambda},\end{equation}

Thus, even in the linear approximation used in [1], we do not get
the correct spin algebra.  Moreover, from the
above algebra it is tempting to absorb the overall factor
$\,(\pi^2 - e\tilde F\cdot S)^{-1}(-e\tilde F\cdot S)\,$ in a redefinition
of the spin operators, but it is clear that the new operator would
be meaningless in the  free limit ($e\to 0$).
Note that in the non-interacting limit
($e\to 0$) we recover (1) as expected.
As in the free case
we have nonvanishing crossed commutation relations between intrinsic
and orbital variables:

\begin{eqnarray}\left\lbrack n_{\beta}\,,\,\pi^{\alpha}\right\rbrack &=&
-\,{\frac {i\,e\,F^{\alpha\gamma}n_{\gamma}\pi_{\beta}}{\G}}\,,\,
\;\lbrack p^{(n)}_{\alpha},\,\pi_{\nu}\rbrack \;= -\;{\frac
{i\,e\,\pi_{\alpha}F_{\nu\gamma}p^{(n)\gamma}}{\G}}\quad\\
\lbrack x_{\mu}\;,\,\pe_{\nu}\rbrack &=&\;{\frac {i\,\pe_{\mu}
\pi_{\nu}}{\G}}\;,\quad\left\lbrack x_{\mu}\;,\;n_{\nu}\right\rbrack\;=\;
{\frac {i\,n_{\mu}\pi_{\nu}}{\G}}\quad . \end{eqnarray}

Therefore the interacting model proposed in [1] suffers from the
same problems of the free one.

Finally, it is important to observe that
the six degrees of freedom represented
by the canonical variables $\,(n_{\mu},\pe_{\nu})\,$, introduced in
[1] to describe spin, should be eliminated by the
constraints. However, in [1] there is just
one first class constraint ($\,\Phi_2 \approx 0\,$)
and three second class ones ($\,\varphi_1=0,\,\varphi_2=0\,$ and
$\,n^2 + 1=0\,$) on the variables $\,(n_{\mu},\pe_{\nu})\,$,
thus we are left with one exceeding degree of freedom. This problem
is related to another one, namely, one can easily verify that the
basic brackets (3) of [1] do not satisfy the Jacobi identity
$\,\left\{p^{(n)\alpha},\left\{ p^{(n)\beta},n^{\gamma}
\right\}\right\} +\,$
cyclic $\,= g^{\alpha\gamma}n^{\beta} - g^{\beta\gamma}n^{\alpha} \ne 0\,$.
Both problems can be solved by introducing
another strong condition, e.g.,
$\,\pe\cdot n =0 \,$ , besides $\,n^2 + 1=0\,$,and defining the
the corresponding Dirac brackets. In this case one recovers
the model defined in (2.12-14) of Plyushchay's work [2] which leads also
to the incorrect spin algebra (1).

\bigskip

This work is partially supported by CNPq and Fapesp.

\bigskip

\noindent [1] M. Chaichian, R. Gonzalez Felipe and D. Louis Martinez,
Phys. Rev. Lett. {\bf 71}, 3405 (1993).

\noindent [2] M. S. Plyushchay, Int. Journal of Mod. Phys. {\bf A7}, 7045
(1992).

\end{document}